\input harvmac 
\input epsf 
%
\noblackbox 
\def\npb#1#2#3{{\it Nucl.\ Phys.} {\bf B#1} (19#2) #3} 
\def\plb#1#2#3{{\it Phys.\ Lett.} {\bf B#1} (19#2) #3} 
\def\prl#1#2#3{{\it Phys.\ Rev.\ Lett.} {\bf #1} (19#2) #3}

\def\atmp#1#2#3{{\it Adv.\ Theor.\ Math.\ Phys.} {\bf #1} (19#2) #3} 
\def\jhep#1#2#3{{\it JHEP\/} {\bf #1} (19#2) #3} 
\newcount\figno 
\figno=0 
\def\fig#1#2#3{ 
\par\begingroup\parindent=0pt\leftskip=1cm\rightskip=1cm\parindent=0pt 
\baselineskip=11pt 
\global\advance\figno by 1 
\midinsert 
\epsfxsize=#3 
\centerline{\epsfbox{#2}} 
\vskip 12pt 
{\bf Fig.\ \the\figno: } #1\par 
\endinsert\endgroup\par 
} 
\def\figlabel#1{\xdef#1{\the\figno}} 
\def\encadremath#1{\vbox{\hrule\hbox{\vrule\kern8pt\vbox{\kern8pt 
\hbox{$\displaystyle #1$}\kern8pt} 
\kern8pt\vrule}\hrule}} 
 
\def\frac#1#2{{#1 \over #2}}

\def\semi{\subset\kern-1em\times\;} 
\def\bar#1{\overline{#1}} 

\def\CO{{\cal O}}

\def\C{{\bf C}}

\def\pa{2 \pi \alpha'} 
\def\One{{1\hskip -3pt {\rm l}}}     
\Title{\vbox{\baselineskip12pt 
\hbox{hep-th/0005031} 
\hbox{EFI-2000-15} 
\vskip-.5in}} 
{\vbox{\centerline{D-branes and Strings as Non-commutative Solitons}  
\bigskip}} 
\medskip\bigskip 
\centerline{Jeffrey A. Harvey, Per Kraus, Finn Larsen and Emil J. Martinec} 
\bigskip\medskip 
\centerline{\it Enrico Fermi Institute and Department of Physics} 
\centerline{\it  University of Chicago,  
Chicago, IL 60637, USA} 
\baselineskip18pt 
\medskip\bigskip\medskip\bigskip\medskip 
\baselineskip16pt 
 
The non-commutative geometry of a large auxiliary $B$-field 
simplifies the construction of D-branes  
as solitons in open string field theory. 
Similarly, fundamental strings are constructed as  
localized flux tubes in the string field theory. 
Tensions are determined exactly using general  
properties of non-BPS branes, and   
the non-Abelian structure of gauge fields on  
coincident D-branes is recovered.   
\Date{May, 2000} 
\lref\wittenflux{E.~Witten,
``Theta Vacua In Two-Dimensional Quantum Chromodynamics,''
Nuovo Cim.\  {\bf A51}, 325 (1979).}
\lref\bsz{N.~Berkovits, A.~Sen and B.~Zwiebach, 
``Tachyon condensation in superstring field theory,'', 
JHEP {\bf 0003}:002, 2000;
hep-th/0002211.} 
\lref\sena{A.~Sen, ``Stable non-BPS bound states of BPS D-branes,'' 
\jhep{9808}{98}{010}, hep-th/9805019;  
``SO(32) spinors of type I and other solitons on brane-antibrane pair,'' 
\jhep{9809}{98}{023}, hep-th/9808141; 
``Type I D-particle and its interactions,'' 
\jhep{9810}{98}{021}, hep-th/9809111;  
``Non-BPS states and branes in string theory,''  
hep-th/9904207, and references therein.} 
\lref\sennon{A. Sen, ``BPS D-branes on non-supersymmetric cycles,''  
\jhep{9812}{98}{021}, hep-th/9812031.} 
\lref\bergman{O.~Bergman and M.~R.~Gaberdiel, 
``Stable non-BPS D-particles,'' \plb{441}{98}{133}, hep-th/9806155.} 
\lref\senuniv{A.~Sen, 
``Universality of the tachyon potential,'' 
JHEP {\bf 9912} (1999) 027, hep-th/9911116.} 
\lref\polchinski{J.~Polchinski, ``Dirichlet-Branes and Ramond-Ramond  
Charges,'' \prl{75}{95}{4724}, hep-th/9510017.} 
\lref\senspinors{A.~Sen, ``$SO(32)$ Spinors of Type I and Other Solitons on  
Brane-Antibrane Pair,'' \jhep{9809}{98}{023}, hep-th/9808141.} 
\lref\ewk{E. Witten, ``D-Branes and K-Theory,'' \jhep{9812}{98}{019};  
hep-th/9810188.} 
\lref\phk{P. Ho\v rava, ``Type IIA D-Branes, K-Theory, and Matrix Theory,''  
\atmp{2}{99}{1373}, hep-th/9812135.} 
\lref\mm{R. Minasian and G. Moore, ``K-Theory and Ramond-Ramond Charge,''  
\jhep{9711}{97}{002}, hep-th/9710230.} 
\lref\yi{P. Yi, ``Membranes from Five-Branes and Fundamental Strings from 
D$p$-Branes,'' \npb{550}{99}{214}; hep-th/9901159.} 
\lref\senpuz{A. Sen, ``Supersymmetric World-volume Action for Non-BPS 
D-branes,'' \hfill\break 
JHEP {\bf 9910}:008, 1999; hep-th/9909062.} 
\lref\senzw{A. Sen and B. Zwiebach, ``Tachyon Condensation in String Field  
Theory,'' hep-th/9912249.} 
\lref\bcr{M. Bill\'o, B. Craps and F.Roose, ``Ramond-Ramond couplings of 
non-BPS D-branes,'' \jhep{9906}{99}{033}; hep-th/9905157.} 
\lref\gms{R. Gopakumar, S.Minwalla and A.Strominger, ``Noncommutative  
Solitons,'' hep-th/0003160.} 
\lref\sw{N. Seiberg and E. Witten, ``String Theory and Noncommutative  
Geometry,'' JHEP {\bf 9909}:032, 1999; hep-th/9908142.} 
\lref\hk{J. A. Harvey and P. Kraus, ``D-Branes as Lumps in Bosonic 
Open String Field Theory,'' JHEP {\bf 0004}:012,2000, hep-th/0002117.} 
\lref\hkm{J. A. Harvey, D. Kutasov and E. Martinec, ``On the Relevance of 
Tachyons,'' hep-th/0003101.} 
\lref\ks{J. Kogut and L. Susskind ``Vacuum Polarization and the  
Absence of Free Quarks in Four-Dimensions,'' Phys. Rev. {\bf D9} 
(1974) 3501.}  
\lref\cds{A.Connes, M. R. Douglas and A.Schwarz, ``Noncommutative Geometry 
and Matrix Theory:Compactification on Tori,'' JHEP {\bf 9802}, 003 (1998), 
hep-th/9711162.} 
\lref\sendes{A. Sen, ``Descent Relations Among Bosonic D-branes,'' 
Int.J. Mod. Phys. {\bf A14} (1999) 4061, hep-th/9902105.} 
\lref\kjmt{R. de Mello Koch, A. Jevicki, M. Mihailescu and R.Tatar, 
``Lumps and P-Branes in Open String Field Theory,'' hep-th/0003031.} 
\lref\bhy{O.Bergman, K.Hori and P.Yi, ``Confinement on the Brane,'' 
hep-th/0002223.} 
\lref\kosts{V. A. Kostelecky and S.Samuel, ``On a Nonperturbative Vacuum for  
the Open Bosonic String,'' Nucl.Phys. {\bf B336} (1990) 263.} 
\lref\mtaylor{N. Moeller and W. Taylor, ``Level Truncation and 
the Tachyon in Open Bosonic String Field Theory,'' hep-th/0002237.} 
\lref\wtaylor{W. Taylor, ``D-brane Effective Field Theory From String 
Field Theory,'' hep-th/0001201.} 
\lref\hkm{J. A. Harvey, D. Kutasov and E. J. Martinec, 
``On the relevance of tachyons'', hep-th/0003101.} 
\lref\bfss{T. Banks, W. Fischler, S.H. Shenker and L. Susskind, 
``M Theory As A Matrix Model: A Conjecture'',  
Phys.Rev.{\bf D55} (1997) 5112, hep-th/9610043.} 
\lref\dwhn{B.~de Wit, J.~Hoppe and H.~Nicolai, 
``On the quantum mechanics of supermembranes'', 
Nucl.\ Phys.\  {\bf B305} (1988) 545.} 
\lref\garousi{M.~R.~Garousi, 
``Tachyon couplings on non-BPS D-branes and Dirac-Born-Infeld action,'' 
hep-th/0003122.} 
\lref\senact{ 
A.~Sen, 
``Supersymmetric world-volume action for non-BPS D-branes,'' 
JHEP {\bf 9910} (1999) 008, 
hep-th/9909062.} 
\lref\bergs{ 
E.~A.~Bergshoeff, M.~de Roo, T.~C.~de Wit, E.~Eyras and S.~Panda, 
``T-duality and actions for non-BPS D-branes,'' 
hep-th/0003221.} 
\lref\kluson{ 
J.~Kluson, 
``Proposal for non-BPS D-brane action,'' 
hep-th/0004106.} 
\lref\dvv{
L.~Motl,
``Proposals on nonperturbative superstring interactions''
hep-th/9701025;
T.~Banks and N.~Seiberg,
``Strings from matrices,''
Nucl.\ Phys.\  {\bf B497} (1997) 41,
hep-th/9702187;
R. Dijkgraaf, E. Verlinde and H. Verlinde, 
``Matrix String Theory'', 
Nucl.Phys. {\bf B500} (1997) 43-61, 
hep-th/9703030.} 
\lref\callm{ 
C.~G.~Callan and J.~Maldacena,  
``Brane Death and Dynamics from the Born-Infeld Action,'' 
Nucl.Phys. {\bf B513} (1998) 198, 
hep-th/9708147.} 
\lref\gibb{G.~W.~Gibbons, 
``Born-Infeld Particles and Dirichlet P-Branes'' 
Nucl.Phys. {\bf B514} (1998) 603, 
hep-th/9709027.} 
\lref\hhk{J.~A.~Harvey, P. Kraus and P. Ho\v rava, 
``D Sphalerons and the Topology of String Configuration Space,'' 
JHEP {\bf 0003}:021,2000, 
hep-th/0001143.}  
\lref\corns{L.Cornalba and R.Schiappa, hep-th/9907211.} 
\lref\ishi{N. Ishibashi, ``A Relation Between Commutative and 
Noncommutative Descriptions of D-Branes,'' hep-th/9909176.} 
\lref\lwone{ 
F.~Larsen and F.~Wilczek, 
``Classical Hair in String Theory I: General Formulation,'' 
Nucl.\ Phys.\  {\bf B475} (1996) 627. 
{hep-th/9604134}.} 
\lref\hlw{P. S.Howe, N.D.Lambert and P.C. West, ``The Selfdual String
Soliton,'' Nucl.Phys. {\bf B515} (1998) 203, hep-th/9709014.}
\lref\dmr{K. Dasgupta, S. Mukhi, and G. Rajesh,
``Noncommutative Tachyons'', hep-th/0005006.}
\lref\malda{J.~Maldacena,
``The large N limit of superconformal field theories and supergravity,''
Adv.\ Theor.\ Math.\ Phys.\  {\bf 2} (1998) 231, hep-th/9711200.}
\lref\luroy{J.X. Lu and S. Roy,
``(p + 1)-Dimensional Noncommutative Yang-Mills and D($p - 2$) Branes'',
hep-th/9912165.}
\lref\sbz{Nathan Berkovits, Ashoke Sen and Barton Zwiebach,
``Tachyon Condensation in Superstring Field Theory'',
hep-th/0002211.}
\lref\wadia{A.~Dhar, G.~Mandal and S.~R.~Wadia,
``String field theory of two-dimensional QCD: 
A Realization of $W_\infty$ algebra,''
Phys.\ Lett.\  {\bf B329} (1994) 15, hep-th/9403050;
``Nonrelativistic fermions, coadjoint orbits of $W_\infty$
and string field theory at $c=1$,''
Mod.\ Phys.\ Lett.\  {\bf A7} (1992) 3129,
hep-th/9207011.}

 
\newsec{Introduction} 
 
General arguments \refs{\sendes,\senuniv,\senzw}, 
explicit calculations  in truncated  open string 
field theory \refs{\hk,\kjmt,\kosts,\sbz}, and renormalization group 
analysis of relevant boundary perturbations \hkm\ all 
suggest that D-branes can be constructed as solitons or lumps 
in open string field theory. Analogies between tachyon condensation 
in open string theory and confinement of electric charge \senpuz\  
have also motivated suggestions that macroscopic closed strings can be  
described by flux tubes in open string field theory \refs{\hk, \bhy, \yi}. 
It is known that fundamental strings ending on D-branes 
can be viewed as flux tubes \refs{\callm,\hlw, \gibb}, the new element 
is the idea that such a description should be valid even in 
the absence of D-branes.  
 
One problem in trying to construct these solutions 
explicitly is that they are string scale objects, and the  
string field theory action contains an infinite number of higher  
derivatives with coefficients set by the string scale. As a result it 
is difficult to obtain an accurate description of $Dp$-branes for 
small $p$, the non-Abelian structure on multiple D-branes is 
far from obvious, and it has not been possible to obtain a 
concrete understanding of the flux tube solution.  
 
In this paper we will show that these difficulties can be 
overcome  by the introduction of 
non-commutative geometry via a background 
$B$-field \refs{\cds, \sw}.  
We consider solitons on the world-volume of an 
unstable D-brane in the presence of a large $B$-field.  The solutions we 
study are the non-commutative solitons recently constructed by 
Gopakumar, Minwalla and Strominger (GMS) \gms\ .  The solitons,  
though string scale with respect to the original space-time 
metric, are much larger than the string scale as measured by the effective 
open string metric of Seiberg and Witten \sw; it is this latter  
fact which facilitates the analysis. Far from the solitons the tachyon  
on the original unstable D-brane will have condensed to its local  
minimum, so that the soliton field configuration 
is asymptotic to the closed string  
vacuum without D-branes. The $B$-field is thus pure gauge far from the  
solitons, and the solutions represent D-branes and strings in the  
standard closed string vacuum. 
 
A remarkable feature of the GMS solitons is that many of their properties 
are insensitive to details of the field theory to which they are  
solutions --- in our case open string field theory on the unstable 
D-brane.  Thus, for the most part we will consider only 
the low-energy dynamics of the light modes of the open string field. 
For most of our considerations we do not need to know the detailed shape of 
the tachyon potential,  we require only the height of 
its local maximum.  
Happily, this is one of the few properties of 
unstable D-branes that we understand precisely, from a conjecture by  
Sen \refs{\sendes,\sena}.  
Combining these observations therefore allows us to verify  
that the solitons have exactly the right tensions to be identified as  
D-branes. We also obtain the correct content of  world-volume fields  
on the D-branes, and find the expected non-Abelian gauge structure  
for multiple D-branes.    
 
Classical open string field theory does not contain closed strings, 
they appear only as poles in loop diagrams. The arguments referred 
to earlier suggest however that after tachyon 
condensation the open string degrees of freedom are frozen out and 
one should find macroscopic  
closed strings in the classical theory.  
In the semi-classical limit these should be  
interpretable as solitons. Indeed, we find that the magic of  
non-commutativity facilitates a concrete construction where 
the fundamental string is identified with a flux tube. The 
fluctuations  of the flux tube can be analyzed explicitly and, 
in a suitable approximation, are described by the  
Nambu-Goto action.  
 
The paper is organized as follows. In section 2 we first review a few  
properties of non-commutative field theories and their solitons, we show 
how to embed this discussion in string theory via unstable D-branes, and 
then proceed to compute the tension of these solitons  in string 
theory  and identify them 
with D-branes. In section 3 we consider the massless gauge fields  
in the string field theory and see how these descend to 
the soliton world-volume.  The gauge fields  
and their interplay with the tachyon field play an important role 
in obtaining the correct D-brane collective dynamics.  
In section 4 we extend the discussion to D-branes 
in type II string theory.
Finally, in section 5, we use similar methods to 
construct a flux tube in the open string field theory.  
We compute its tension and find its excitations, and 
are led to conclude that it can be identified  
with the closed fundamental string expected in the vacuum  
after tachyon condensation. 
 
Notation: we will denote the complete set of space-time coordinates by $x^\mu$, 
non-commutative directions by $x^i$, and the remaining directions by $x^a$. 
 
 
\newsec{Soliton solutions and their tensions} 
 
\subsec{The Non-commutative Limit} 
We will be considering Minkowski space-time with closed 
string metric  $g_{\mu\nu} = \eta_{\mu\nu}$ in  
the presence of a constant $B_{\mu\nu}$ field, the latter taking non-vanishing 
values only in purely spatial directions.   
As explained in \sw, one should distinguish between the closed string metric 
$g_{\mu\nu}$ and the open string metric $G_{\mu\nu}$ which are related by 
\eqn\one{ 
G_{\mu\nu} = g_{\mu\nu} - (\pa)^2(Bg^{-1}B)_{\mu\nu}. 
} 
The open  string field theory action is related to the one without a $B$-field 
by using the metric $G_{\mu\nu}$, replacing ordinary products of fields by  
$\star$ 
products, 
\eqn\oneaa{A(x)B(x) \rightarrow  
A \star B = e^{{i \over 2}\theta^{\mu\nu}  
\partial_\mu \partial_{\nu'}} A(x) B(x')|_{x=x'}~,} 
and replacing the open string coupling $g_s$ by 
\eqn\onea{G_s = g_s \biggl( {\det G \over \det(g + \pa B) } \biggr)^{1/2}.} 
Here  
\eqn\two{ 
\theta^{\mu\nu}=-(\pa)^2 \left({1 \over g + \pa B}B 
{1\over g - \pa B}\right)^{\mu\nu}.  
} 

Now there is a new dimensionless parameter $\alpha' B_{\mu\nu}$,  
or equivalently $\theta^{\mu\nu}/\alpha'$, 
that can be varied in order to simplify the analysis. Denoting the directions 
in which $B_{\mu\nu}$ is non-vanishing as $x^i$, we 
will be  interested in the limit of large non-commutativity,   
$\alpha' B_{ij} \rightarrow \infty$ with $g_{ij}$ held fixed.  
In this limit the  solitons will 
become much larger than the string scale when measured in the open 
string metric and this will lead to many simplifications.  To avoid confusion, 
we note that there is  an equivalent, but 
perhaps more familiar, form 
of the limit: $\alpha' B_{ij} \rightarrow 0$, $g_{ij} \rightarrow 0$, 
and $G_{ij}$ is fixed.   
In this form one has $\theta^{ij}/\alpha' \rightarrow \infty$. 
The two versions of the limit are simply related by a coordinate  
transformation, 
$x^{i} \rightarrow 2\pi \alpha' B_{ij} x^j$.  In either form of the limit,  
\eqn\thlim{\theta^{ij} = \left( {1 \over B}\right)^{ij}.} 

\subsec{Solitons in String Field Theory} 
 
Let us review some aspects of D-branes as solitons in open string 
field theory.  Our primary focus will be solitons on the world-volume of  
a bosonic D25-brane, although we will also discuss type II D-branes.   
 
The bosonic D25-brane has on its world-volume a tachyon, $m_t^2 = -1/\alpha'$, 
with a potential indicated schematically  in fig. 1. 
We keep an explicit factor of the D25-brane tension, $T_{25}$, 
in front of the action, so that the physical tachyon potential is  
$T_{25} V(t)$, and we have shifted the tachyon so that the local minimum is 
at $t=0$. The unstable local maximum $t=t_*$ represents the space filling 
D25-brane, with $T_{25}V(t_*) = T_{25}$.  The local minimum $t=0$ is the 
closed string vacuum without D-branes, $V(0)=0$.   
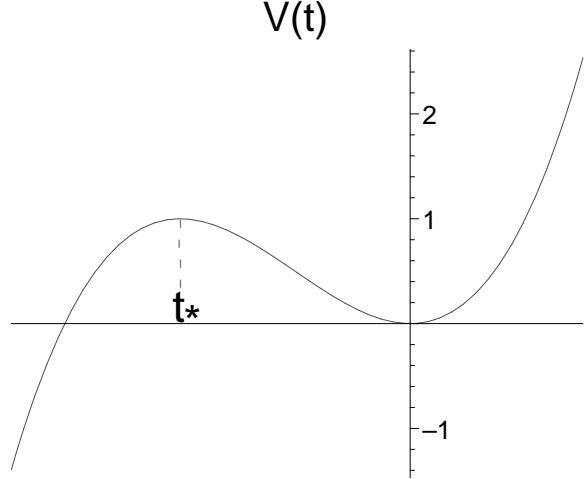
\fig{The bosonic open string tachyon potential.}{tach.ps}{3truein} 
 
The tachyon action supports unstable soliton solutions which are 
asymptotic to the closed string vacuum at $t=0$, 
and it has been proposed to identify these with bosonic Dp-branes with 
$p<25$.  Such solitons have been constructed numerically in level truncated 
open string field theory in \refs{\hk,\kjmt}, and for sufficiently large $p$ 
good agreement was found between the tension and low lying spectra of the 
soliton and those of bosonic Dp-branes.  However, the presence of higher 
derivatives in the string field theory action greatly complicates the analysis, 
and it seems very challenging to recover such fundamental properties as  
enhanced gauge symmetry for coincident D-branes.   
 
In the present work we instead apply some powerful simplifications following 
from non-commutative geometry.   
Our starting point is an effective  action for the tachyon obtained by  
integrating out (classically) 
all fields in the string field theory action which are ``sourced'' 
by the tachyon,  
\eqn\five{ 
S= {C \over g_s}\int \! d^{26}x\,  
\sqrt{g}\left({1 \over 2}f(t)g^{\mu \nu}\partial_\mu t  
\partial_\nu t + \cdots  - V(t)\right)~, 
} 
where $\cdots$ indicate higher derivative terms which will not be  
written explicitly, and we have explicitly 
displayed the string coupling by defining   
a $g_s$-independent constant 
\eqn\C{C = g_s T_{25}~.} 
According to the conjecture of \senuniv\ the entire  
action vanishes at the local minimum: $f(0) = V(0) = 0$  
(with corresponding equations for the higher derivative terms).   
 
Now we turn on the $B$-field, which changes the action to 
\eqn\fiveG{ 
S= {C \over G_s}\int \! d^{26}x \, 
\sqrt{G}\left({1 \over 2}f(t)G^{\mu \nu}\partial_\mu t  
\partial_\nu t + \cdots  - V(t)\right)~, 
} 
where $\star$ products are now implied.  Due to the non-commutativity one needs 
to specify an ordering of fields to define \fiveG, but this level of precision 
will not be needed for the present analysis. 
 
Soliton solutions in theories of this kind were constructed in \gms\ in the 
limit of large non-commutativity.  A simple scaling computation shows that the  
potential term dominates over the derivative terms in this limit, so  
that the equation of motion for static solitons is 
\eqn\soleq{{dV \over dt} = 0~.} 
Localized soliton solutions to this equation exist due to the presence of  
the $\star$ product. The construction of GMS relies on the existence of  
functions $\phi$ satisfying\foot{Such functions have also made
an appearance in the work of \wadia.}
\eqn\phieq{ 
\phi \star \phi = \phi~, 
} 
since then 
\eqn\bla{F(\lambda \phi) = F(\lambda)\phi~,} 
for any function $F$ of the form $F(x)= \sum_{n=1}^\infty a_n x^n$.   
In particular, 
\eqn\deriv{{dV \over dt}|_{t=\lambda\phi} = 
\left({dV \over dt}|_{t=\lambda}\right) \phi~,} 
and \soleq\ is solved by choosing $\lambda$ to be an extremum of $V$. 
Turning on $B_{ij}$ in two directions, say $x_{1,2}$,  the 
simplest function satisfying \phieq\ is the Gaussian 
\eqn\phisol{\phi_0(r) = 2 e^{-r^2/\theta}, \quad r^2 = x_1^2 +x_2^2~,} 
with $B=B_{12}$, $\theta = 1 / B$. For the potential indicated  
in fig. 1 the solution will thus be  
\eqn\soliton{t= t_* \phi_0(r)~.} 
Note that for this solution the tachyon asymptotically approaches its value 
$t=0$ in the closed string vacuum.  The resulting object is a 23+1 dimensional 
soliton that we will identify with the D23-brane.  
The coordinate size of the soliton 
is $\Delta x \approx \sqrt{\theta}= 1/\sqrt{B}$,  
which goes to zero in the large B 
limit.  However, for determining the importance of $\alpha'$ corrections 
the relevant quantity is $\Delta x_{{\rm open}} = 
\sqrt{G_{ij} \Delta x^i \Delta x^j} \approx  {\alpha' \sqrt{B}}$. 
In the limit $\alpha'B \rightarrow \infty$  
this is much larger than $\sqrt{\alpha'}$, so $\alpha'$ 
corrections, in the form of the derivative terms in \fiveG, are suppressed. 
 
The above construction easily generalizes to arbitrary even codimension 
solitons, for example by turning on equal $B$-fields in the $(12),(34) \ldots 
(2q-1,2q)$ 
planes and  by replacing $r^2$ in \phisol\ by $r^2 = x_1^2 +x_2^2 
+ \cdots + x_{2q}^2$.  The resulting soliton is to be identified with a 
D($25-2q$)-brane. 
 
\subsec{Tension of solitons} 
 
We now show that our solutions have the same tension as bosonic Dp-branes, 
\eqn\ten{ 
T_p = (2\pi)^{25-p}(\alpha')^{(25-p)/2}T_{25}~. 
} 
We first consider the D23-brane soliton.  At large non-commutativity we 
neglect the explicit transverse derivatives in \fiveG , and the action for  
translationally invariant configurations along the D23-brane is 
\eqn\sv{S=-{C \over G_s} \int \! d^{26}x \, \sqrt{G}V(t)~.} 
Now we insert the soliton solution, use $V(t) = V(t_*)\phi_0(r)$, and 
integrate over $x_1, x_2$: 
\eqn\comten{S = - {C V(t_*) \over G_s} \int \! d^{24}x \int \! d^2{x}\, 
\sqrt{G} \phi_0(r) =  
- {2 \pi \theta C V(t_*) \over G_s} \int \! d^{24}x \, \sqrt{G}~.} 
Next 
we use the relation \onea\ between $G_s$ and $g_s$, which for large   
$B$-field is 
\eqn\Gg{G_s = {g_s \sqrt{G} \over 2 \pi \alpha' B \sqrt{g}}~.} 
In our conventions $V(t_*)=1$ so, 
inserting this into \comten\ and using $\theta = 1/B$, we find 
\eqn\tenres{S = - (2\pi)^2 \alpha' {C \over g_s} \int \! d^{24}x \, \sqrt{g}~.} 
Finally, recall $C = T_{25}g_s$. This identifies the tension of the  
soliton as  
\eqn\tenresb{T_{{\rm sol}}= (2\pi)^2 \alpha' {C \over g_s} =  
(2\pi)^2 \alpha' T_{25} = T_{23}~.} 

Remarkably, in this limit we get precisely the correct answer without 
knowing  
the detailed form of the tachyon potential.  We only need  
its value at the unstable extremum which follows  from the conjecture  
of Sen, as  substantiated in the work of 
\refs{\kosts,\senzw,\mtaylor,\hkm}.  
It is straightforward to  generalize this result to  arbitrary even codimension 
solitons, and to reproduce the formula \ten\ for odd $p$.

\subsec{Multiple D-branes} 
 
There is a one-to-one correspondence between functions on the 
non-commutative $R^2$ transverse to the D23-brane, thought of as 
the phase space of a particle in one dimension, and operators 
acting on the Hilbert space of one-dimensional particle
quantum mechanics.  Multiplication by 
the $\star$ product goes over to operator multiplication, 
and integration over  $R^2$ corresponds to tracing over the Hilbert space, 
\eqn\corr{ A \star B \leftrightarrow \hat{A} \hat{B}~, \quad\quad 
{1 \over 2\pi \theta} \int \! d^2x_i \leftrightarrow {\rm Tr}~.} 

Under this correspondence, the  equation \phieq\ becomes  
the equation for a projection operator. 
This correspondence was utilized in 
\gms\ to construct more general soliton solutions.   
The soliton solution $t= t_* \phi_0$ \soliton\ corresponds  to the  
projection operator onto the ground state of a one-dimensional 
harmonic oscillator, $\phi_0 \sim |0\rangle\langle 0|$ .  
Other solutions are obtained 
by choosing other projection operators,  $\phi_n \sim |n\rangle\langle n|$, 
or we can choose a superposition 
(a level $k$ solution in the terminology of GMS) 
\eqn\fourteen{ 
t_k = t_*(\phi_0 + \phi_1 + \ldots + \phi_{k-1})~. 
} 
Since the projection operators are orthogonal, the energies just add 
\eqn\fifteen{ 
V(t_k) = kV(t_1)~. 
} 
Thus  this configuration corresponds to $k$ coincident 
D-branes; further evidence for this claim  will appear in  
succeeding sections.   

Since the $\phi_m=|{m}\rangle\langle{m}|$ are a complete set
of projection operators, the limit $k\rightarrow\infty$ of
the level $k$ solution is $t_\infty=t_*\One$.
This solution can be identified with the D25-brane with no
tachyon condensate; indeed, 
the tachyon takes the value $t=t_*$ everywhere, and
the energy density is that of the D25-brane.
At large $k$, the level $k$ solution \fourteen\ 
represents, in the basis constructed in \gms,
a lump of size $r_k\sim\sqrt k$, which approximates the
string field configuration of an unstable D25-brane for smaller radius, 
and the closed string vacuum state for larger radius.
An amusing case is the projection operator complementary
to \soliton, namely $t=t_*(1-\phi_0)$.  
Evaluating the energy of this configuration, 
one formally finds the energy of a D25-brane 
`minus' that of a D23.\foot{Of course,
the energy of the D25 is infinitely larger than that of the
D23; to make this statement precise, one must go to finite 
volume, {\it e.g.} by compactifying the system on a large torus.}
 
In the limit of infinite non-commutativity the action \fiveG\ can be written 
in operator form as 
\eqn\sop{S = T_{23} \int \! d^{24}x_a \,{\rm Tr}\left( {1 \over 2} 
f(\hat{t}) 
G^{ab} \partial_{a}\hat{t} \partial_{b} 
\hat{t} - V(\hat{t}) \right),} 
where $x^{a}$ are the commutative directions.    
 The action in operator form has a manifest  
$U(\infty)$ symmetry  
\eqn\sym{\hat{t} \rightarrow U \hat{t} U^\dagger~.} 
These group operations are familiar from the construction of the Matrix 
theory membrane \refs{\bfss,\dwhn}:  
They correspond to area preserving diffeomorphisms.  
This is no accident.  The Matrix theory membrane in non-compact 
space is essentially a D2-brane that has been bound to an infinite number 
of D0-branes such that the D0 charge density is finite 
(represented by a magnetic flux $F$).  
This charge density is equivalent to the $B$-flux  
of non-commutative geometry, since $B$ and $F$ are indistinguishable 
on the brane.  Thus the two constructions are identical, 
and it is useful to keep this relationship in mind. 
The main difference between the two situations is the presence 
of the tachyon field. 
 
It is clear that, if we neglect the standard kinetic term, then acting with 
an area preserving diffeomorphism on a given configuration is a symmetry 
of \fiveG\ 
since it preserves the $\star$ product (which is defined in terms of the  
volume two-form)  and the integration measure.   On the other hand, the  
standard kinetic term involves also the metric $g_{ij}$.  Only the 
elements of $U(\infty)$ corresponding to translations and rotations 
preserve this metric and so are exact symmetries of the action. 
Thus, for purely scalar actions, as considered by GMS, the $U(\infty)$ 
is an approximate global symmetry that becomes exact in the limit of 
infinite non-commutativity.  
 
Solutions of the form \fourteen\ preserve a $U(k)$ subgroup of the  
$U(\infty)$ global symmetry of the action at infinite $\theta$ 
(times an irrelevant group acting in the orthogonal space), 
and all excitations on the branes will transform in  
multiplets of $U(k)$. The breaking of $U(\infty)$ to $U(k)$ leads to 
an infinite number of Nambu-Goldstone bosons in the spectrum of the soliton.  
Again, for a purely scalar theory, these become pseudo-Nambu-Goldstone 
bosons at finite $\theta$.  As we discuss in section 3, the story 
changes after coupling to the gauge fields on the D-brane. 
 
\subsec{Tachyon on the soliton} 
The D25-brane tachyon reflects the fact that  
$t=t_*$ is an unstable point of the potential: 
\eqn\eleven{ 
V(t) = V(t_*) + {1 \over 2}  V''(t_*) (t-t_*)^2 + \cdots~, \quad V''(t_*)<0, 
} 
and from \five\ the mass of the open string tachyon is 
\eqn\tm{m_t^2={ V''(t_*) \over f(t_*)} = - {1 \over \alpha'}~.} 
The lower D-branes also have a tachyon with this mass, and we should recover 
it by studying fluctuations around the soliton. 
 
As usual, we first consider the simplest case of two non-commutative directions 
$x^i$. Using the operator correspondence, a complete set of functions on 
this space is given by 
\eqn\comp{\phi_{mn}(x^i) \sim |m \rangle \langle n|~,} 
so the general fluctuation is 
\eqn\genfluc{t +\delta t(x^\mu) = t_* \phi_{00}(x^i) +  
\sum_{m,n=0}^\infty \delta t_{mn}(x^a)\phi_{mn}(x^i)~.} 
Reality of $t$ requires  $\delta t_{mn}$ to be a Hermitian matrix. 
As we have discussed, the fluctuations include Nambu-Goldstone bosons 
from the spontaneous breaking of $U(\infty)$ by the soliton.  As in GMS, 
the generators of $U(\infty)$ are  
$R_{mn} = |m\rangle \langle n| +  |n\rangle \langle m|$ and 
$S_{mn} = i(|m\rangle \langle n| -  |n\rangle \langle m|)$, and the  
Nambu-Goldstone bosons are the nonzero components of $\delta t = 
[R_{mn},t],[S_{mn},t]$.  We will see in the next section that   
in the D-brane application of interest $U(\infty)$ is  a  
gauge symmetry and the Nambu-Goldstone bosons are eaten by the Higgs 
mechanism.  Thus we use the $U(\infty)$ symmetry to set 
\eqn\uni{\delta t_{0m} = \delta t_{m0} = 0~, \quad m \ge 1~.} 

Substituting into the action and using the fact that 
$\phi_0 \equiv \phi_{00}$ is orthogonal to 
$\phi_{mn}$ for $m,n >1$ we are left with 
the single tachyon fluctuation $\delta t_{00}$ which we rename $\delta t$. 
We now use   
\eqn\twelve{ 
V(t+ \delta t\phi_0) = V(t_* + \delta t)\phi_0 =  
\left(V(t_*) + {1 \over 2} V''(t_*)(\delta t)^2 +\ldots\right) \phi_0~. 
} 
After integrating over the soliton the 
quadratic effective action for $\delta t$ becomes 
\eqn\thirteen{ 
S = T_{23} \int \! d^{24}x_a \, \left( {1 \over 2}  
f(t_*)\partial^{a} \delta t \partial_{a} \delta t 
- {1 \over 2} V''(t_*) (\delta t)^2 \right)~, 
} 
so the D23-brane correctly inherits the tachyon of the D25-brane, 
\eqn\tmb{m_t^2={ V''(t_*) \over f(t_*)} = - {1 \over \alpha'}.} 

This analysis can be easily repeated to obtain a tachyon on the 
more general solution \fourteen; then acting with $U(k)$ on this  
tachyon produces the expected $k^2$ tachyons.  A notable point is that 
the general tachyon will contain operators of the form  
$|m\rangle\langle n|$, $m \neq n$. 
These correspond to non-spherically symmetric tachyon fluctuations in the 
transverse space.  So the full $k^2$ of tachyons comes from including both 
spherically symmetric and non-symmetric tachyon configurations.   

 
\newsec{Coupling to Gauge Fields} 
 
A characteristic property of D-branes is the presence of gauge fields on their  
world-volume.  In this section, we demonstrate that there exists a single  
massless gauge field on the non-commutative soliton,  
therefore providing further evidence  
that the soliton  can be identified as a D-brane.  
Furthermore, when we generalize to level $k$ solutions such 
as \fourteen, the gauge symmetry is enhanced to $U(k)$ 
in the appropriate way, with the components of the gauge fields  
transverse to the soliton behaving as the standard 
adjoint Higgs fields on a D-brane. 
Finally, the gauge symmetry removes 
from the spectrum the unwanted (pseudo) Nambu-Goldstone bosons  
describing soft deformations \sym\ of the soliton \soliton; they 
are eaten by the Higgs mechanism. 
 
This last point brings out an intriguing aspect of our situation. 
Ordinarily, when a global symmetry is explicitly broken, 
no matter how softly, it cannot be gauged.  One might then 
wonder how the $U(\infty)$ symmetry \sym\ can be gauged, 
since it is broken by the tachyon kinetic term 
at finite $\theta$. 
The resolution of this puzzle is strikingly reminiscent of 
general relativity.  There, the potential energy term in 
the action of a scalar field 
is invariant under volume-preserving diffeomorphisms; but 
the kinetic energy term is not. 
Rather, the coordinate transformations are gauged by coupling to 
a metric, without there being a corresponding global symmetry 
in the absence of gauging.  We will see that, once we include the  
gauge fields, the approximate $U(\infty)$ global symmetry 
is realized as an exact gauge symmetry.  
In the present context, the non-commutative D-brane  
gauge field takes over the role of the metric in covariantizing 
area-preserving diffeomorphisms! 
 
A note on conventions: 
In this section we often set $2\pi \alpha' =1$ to avoid clutter. 
 
\subsec{The Gauge Theory and its Symmetries} 
First consider the action for the D25-brane.   
Imagine integrating out all fields except  
for the tachyon and the gauge field.   
The result will be some gauge invariant expression  
involving an infinite number of derivatives and an infinite number of  
higher powers of  
the fields.   Working in terms of the 
$\star$ product implies that the gauge transformation law of the  
non-commutative $U(1)$ gauge field is 
\eqn\gt{\delta_\lambda A_{\mu} = \partial_\mu \lambda - i[A_\mu,\lambda],} 
where 
\eqn\comm{ [A_\mu,\lambda] = A_\mu \star \lambda - \lambda \star A_\mu.} 
The corresponding field strength is 
\eqn\fs{F_{\mu\nu} = \partial_\mu A_\nu - \partial_\nu A_\mu -  
i[A_\mu,A_\nu]~.} 
We also need to know the gauge transformation of the tachyon.  String theory 
disk diagrams 
 reveal that the tachyon transforms in the adjoint of the  
non-commutative $U(1)$ \garousi , {\it i.e.} 
\eqn\gttachy{\delta_\lambda t = -i[t,\lambda]~,} 
with corresponding covariant derivative 
\eqn\cd{D_\mu t = \partial_\mu t -i[A_\mu,t]~.} 

We will  restrict attention 
to terms quadratic in the field strength.  The full action contains higher 
derivative terms in the commutative directions, but for clarity's sake these 
will not be displayed explicitly.  The action is then, 
\eqn\tachgauge{S = {C \over G_s} \int \! d^{26}x \, \sqrt{G} \left( 
{1 \over 2}f(t)D^\mu t D_\mu t - V(t) - {1 \over 4} h(t)F^{\mu\nu}F_{\mu\nu} 
\right).} 
As with the tachyon kinetic term, the gauge kinetic term $h(t)$ vanishes 
at the local minimum $t=0$ according to the conjecture of  
Sen. A simple RG flow argument for this result 
is given in \hkm.

The derivatives in non-commutative directions are all suppressed by $\alpha'$,  
and so can be dropped in the large $B$ limit, both in the action and in the 
gauge transformation laws.   
We again take the $B$-field to be non-vanishing in two directions $x^i$   
and denote the remaining coordinates by $x^a$.    
Expanding out the action in the large $B$ limit yields 
\eqn\sexpand{\eqalign{S = {C \over G_s} \int \! d^{26}x \, \sqrt{G}  
\left( \right. 
&{1 \over 2}f(t)D^a t D_a t - {1 \over 2}h(t)D^a A_i D_a A^i 
- V(t) - {1 \over 2} f(t)[A_i,t][A^i,t]  \cr 
&+ {1 \over 4} h(t)[A_i,A_j][A^i,A^j]  
-{1 \over 4} h(t)F^{ab}F_{ab} \left. \right)  
.\cr}} 
The $A_i$ now appear as scalar fields.  The action in this limit is invariant 
under the gauge transformations 
\eqn\gts{\eqalign{\delta_\lambda t &= -i[t,\lambda]~, \cr 
\delta_\lambda A_i &= -i[A_i,\lambda]~, \cr 
\delta_\lambda A_a &=\partial_a \lambda -i[A_a,\lambda]~. \cr}} 
By passing to the operator description in which  
${1 \over 2\pi \theta} \int d^2 x \rightarrow {\rm Tr}$ 
 and fields are replaced by  
matrix representations of Hilbert space operators, one sees that \sexpand\ 
is a 23+1 dimensional $U(\infty)$ gauge theory coupled to the adjoint scalars 
$t, A_i$.  This 23+1 gauge theory emerges even before considering the 
soliton background. Note that the transformation law of 
the tachyon is just the infinitesimal form of \sym ; the $U(\infty)$  
symmetry is a gauge symmetry, as advertised previously. At finite 
non-commutativity the term $\partial_i\lambda$ in $\delta_\lambda A_i$  
should be restored; the gauge symmetry then remains exact. 
 
Before studying the soliton solution and its fluctuations, we briefly 
digress to explain the emergence of the 23+1 dimensional  
$U(\infty)$ gauge theory from another viewpoint 
(besides the relation to Matrix theory given above), essentially repeating 
comments in \sw.  The open string theory effective action at lowest 
order is given by the path integral on the disk $\Sigma$ 
with boundary conditions 
\eqn\bc{g_{ij}\partial_n x^j +  
 B_{ij} \partial_t x^j|_{\partial \Sigma}=0 
~.} 
For $B=0$ one has Neumann boundary conditions, $\partial_n x^i =0$, while 
for $\alpha' B \rightarrow \infty$ one finds Dirichlet boundary conditions, 
$\partial_t x^i =0$.  In the large $B$ limit one thus finds D23-branes, 
though located at arbitrary transverse positions. In fact, this can be 
thought of as a continuous distribution of D23-branes with density  
proportional to $B$ (see for example \refs{\ishi,\luroy}).  
In the large $B$ limit, the theory governing such 
a system would be a 23+1 dimensional $U(\infty)$ gauge theory, which is 
in accord with the result above. 
 
\subsec{Tachyon - gauge field fluctuations about the soliton} 
 
We now consider fluctuations of the action \sexpand\ around the soliton 
solution 
\eqn\tachsol{t = t_* \phi_{00}~.} 
The soliton breaks the $U(\infty)$ gauge symmetry down to $U(1)$ (times 
the group ``$U(\infty -1)$'' which will play no role in the discussion.)  
Working in unitary gauge we can take the tachyon fluctuations as in  
\genfluc, \uni.  The  fluctuations of the gauge field are 
\eqn\gaugefluc{\eqalign{A_a(x^\mu) &=  
\sum_{m,n=0}^{\infty}A_a^{mn}(x^b) \phi_{mn}(x^i)~, 
\cr 
A_i(x^\mu) &= \sum_{m,n=0}^{\infty}A_i^{mn}(x^a) \phi_{mn}(x^j)~, \cr}} 
where $A_a^{mn}$ and $A_i^{mn}$ are Hermitian as matrices with indices 
$mn$.   
Inserting the fluctuations into the action \sexpand\ we find that all modes 
with $m,n>0$ are projected out by the soliton background, as was seen 
previously for the pure tachyon fluctuations.  Remaining are  
$\delta t_{00}$, $A_i^{00}$, $A_a^{00}$, as well as the $0m$ and $m0$ 
components of $A_i$, $A_a$.  For convenience we drop the explicit $00$ 
indices on the first three fields; after integrating over the transverse 
space their action is found to be  
\eqn\soo{S =T_{23} \int \!  d^{24}x  ~\sqrt{g}\, \left( 
{1 \over 2} f(t_*) \partial^a \delta t \partial_a \delta t - V(t_* + \delta t) 
+ {1 \over 2} h(t_*) \partial^a  A_i \partial_a  A_i 
- {1 \over 4} h(t_*) F^{ab}F_{ab}\right)~,} 
where $F_{ab} = \partial_a A_b - \partial_b A_a$ is the standard  
field strength.  This is precisely the right action to describe the tachyon, 
gauge field, and transverse scalars on a D23-brane.  The transverse  
scalars $A_i$ play the role of translational collective coordinate, and 
are guaranteed to be massless by the original non-commutative $U(1)$  
gauge invariance.   
 
The corresponding story for the $0m$ and $m0$ components of $A_i$, $A_a$ 
involves some additional subtleties.   
For $A_i$ we find the action, 
\eqn\sai{S = T_{23} \sum_{m=1}^\infty  
\int \! d^{24}x \, \sqrt{g}  \left( 
{1 \over 2}h(t_*)\partial^a A_i^{0m} \partial_a A_i^{m0} 
- {1\over 2} t_*^2 f(t_*) A_i^{0m} A_i^{m0} \right).} 
A very similar result holds for  
$A_a$. The second term appears to be 
the standard mass term for W-bosons after gauge symmetry breaking, giving 
a mass 
\eqn\gmass{m_A^2 = \left({t_* \over 2\pi\alpha^\prime}\right)^2 ~.} 
We have inserted for definiteness the perturbative values  
$h(t_*)=(\pa)^2, f(t_*)=1$. 
The appearance of the tachyon VEV $t_*$ 
is interesting because it is the only point in this work  
where we need a quantity that 
is not known exactly. 
It is possible to estimate $t_*$ in the level-truncation 
scheme. This yields an expression  
$t_*^2\sim\alpha^\prime$ with a coefficient that 
rapidly converges to some value of ${\cal O}(1)$ \refs{\kosts,\senzw,\mtaylor}. 
 
Although these states have  mass of order the string scale and are 
thus removed from the low energy spectrum, we 
are naively left with infinitely many massive modes on the soliton, 
degenerate as $\theta\to\infty$ and with a spacing proportional to 
$1/\sqrt{\theta}$  at  finite $\theta$, in clear contradiction with 
the known spectrum of D-branes.  
 
We believe the resolution of this puzzle involves the higher 
derivative terms in the commutative directions and the freezing 
out of the open string degrees of freedom in the closed string 
vacuum. For example, Kostelecky and Samuel showed in \kosts\ 
that non-local terms in the string field theory action which 
appear due to the substitution 
\eqn\tildedef{ 
\phi\to {\tilde \phi}\equiv 
e^{\alpha^{\prime}\ln{3\sqrt{3}/4}~\partial_\mu\partial^{\mu}}\phi
} 
in cubic interaction terms, have the effect of 
modifying the tachyon propagator in the presence of tachyon 
condensation so that there is no physical pole. 

It is reasonable to expect that, in our context,  
the apparent infinite tower of massive gauge fields  
on the soliton is similarly removed. It would be nice to 
justify this expectation with an explicit computation, but we 
have not yet done so.  
 
We should also point out that the higher derivative terms do not affect our 
previous calculations in any substantial way.  By writing out the higher 
derivative terms explicitly and repeating our analysis one sees that the 
D23-brane precisely inherits the higher derivative terms with the same 
coefficients as on the D25-brane.  So for the modes considered previously, 
whatever the higher derivative terms do to the spectrum on the D25-brane, 
they do the same thing on the D23-brane. 
 
\subsec{Multiple D-branes} 
Solitons  with the tachyon profile given in \fourteen\  
are interpreted as $k$ coincident  
D23-branes, and we expect to see the action \soo\ replaced by an action with 
$U(k)$ gauge invariance.  First consider the field strength $F_{ab}$. 
Inserting the expansion \gaugefluc\ into \fs\ we find 
\eqn\fsexp{F_{ab}=F_{ab}^{mn}\phi_{mn},} 
with 
\eqn\fsdef{F_{ab}^{mn}=\partial_a A_b^{mn}-\partial_b A_a^{mn} 
-i[A_a,A_b]^{mn}~.} 
In the latter equation $A_a$ are being multiplied as matrices.  We can now 
work out the final term in \sexpand\ for the background \fourteen\ as 
\eqn\gkt{{C \over G_s} \int \! d^{26}x \sqrt{G} \left( -{1 \over 4} 
h(t_k)F^{ab}F_{ab}\right) = T_{23} \int \! d^{24}x \, \sqrt{g} \left( 
-{1 \over 4} \sum_{m=0}^{k-1}\sum_{n=0}^\infty F^{ab \, mn}F_{ab}^{nm} 
\right).} 
Components of the gauge field $A_a^{mn}$ with $m,n > k-1$ have been  
projected out by the soliton background.  There still remain in the action 
\gkt\ components with $m<k$, $n\ge k$ or $m\ge k$, $n<k$, but we conjecture 
that these states are removed from the spectrum by the  mechanism  
described in the previous subsection.  What  
remains then is precisely the gauge kinetic term for a $U(k)$ gauge theory. 
 
Similarly including $t$ and $A_i$ is straightforward.  Ordinary  
derivatives are replaced by $U(k)$ covariant derivatives, and altogether 
we recover the $U(k)$ gauge invariant version of \soo.   This enhancement 
of the gauge symmetry for coincident solitons, though following rather 
easily from the present formalism, is highly nontrivial from a broader 
field theory vantage point, and provides strong evidence for the  
identification of the solitons with D-branes. 
 
\subsec{Massive modes} 
 
Though we have so far focussed on the tachyon and gauge field fluctuations,
it is not hard to generalize our considerations to show that the solitons
correctly inherit from the D25-brane the entire tower of open string 
states.  Consider for illustration the quadratic terms for some massive
field $\psi$,
\eqn\psiact{S= {C \over G_s} \int \! d^{26}x \, \sqrt{G}\left(
{1 \over 2} f(\Phi)\partial_\mu \psi \partial^\mu \psi -{1 \over 2}
m^2(\Phi)\psi^2 \right),}
where $\Phi$ denotes all fields in the theory.  
The mass of $\psi$ on the D25-brane is determined by evaluating 
$f(\Phi)$ and $m(\Phi)$ at the local maximum $\Phi=\Phi_*$.  
The soliton background
corresponds to $\Phi = \Phi_* \phi_0$, and we consider the fluctuations
$\psi = \sum \psi_{mn}\phi_{mn}$.  As before, $m,n \neq 0$ fluctuations
are projected out; $m=0$, $n>0$ and $m>0$, $n=0$ fluctuations are removed
(we conjecture) by higher derivative terms; so what remains, after 
integrating over the soliton, is
\eqn\psiactb{S= T_{23} \int \! d^{24}x \, \left(
{1 \over 2} f(\Phi_*)\partial_\mu \psi_{00} \partial^\mu \psi_{00} -{1 \over 2}
m^2(\Phi_*)\psi_{00}^2 \right).}
Since what appears in the  action are the functions $f(\Phi)$ and $m(\Phi)$ 
evaluated at the local maximum of the potential, $\psi$ has the same mass
on the soliton as on the original D25-brane.  This shows that the spectrum of
the soliton is inherited from the D25-brane, and so can be identified with
the spectrum of a D23-brane including all massive string states. 

\lref\kw{C. Kennedy and A. Wilkins,``Ramond-Ramond Couplings on
Brane-Antibrane Systems,'' Phys. Lett. {\bf B464} (1999) 206; hep-th/9905195.}
\lref\edtach{E.Witten, ``Noncommutative Tachyons and String
Field Theory,'' hep-th/0006071.}
\lref\quillen{D.Quillen, ``Superconnections and the Chern character,''
Topology, {\bf 24} (1995) 89; N. Berline, E.Getzler and M. Vergne,
{\it Heat Kernels and Dirac Operators}, Springer-Verlag (1991).}

\newsec{Non-commutative solitons on Type II D-branes}

The construction of D-branes as non-commutative solitons in the
bosonic string has an obvious extension to Type II superstring
theory. Type IIA theory contains non-BPS Dp-branes for p odd
and IIB theory has non-BPS Dp-branes for p even
\refs{\sena,\bergman,\sennon,\phk} which have a natural interpretation
as sphalerons \hhk.  Solitons in level truncated open superstring field
theory have been studied in \sbz.  

To be concrete consider the non-BPS space-filling $D9$-brane
of IIA theory. As before, we turn on a $B$-field in two dimensions
and consider the limit of large $\alpha' B$. The analog of
equation \fiveG\ is
\eqn\susyact{{C \over G_s}\int \! d^{10}x\, 
\sqrt{G}\left({1 \over 2}f(t)G^{\mu \nu}\partial_\mu t 
\partial_\nu t + \cdots  - V(t)\right)~,}
with 
\eqn\susyc{C = g_s T_{9A}~.}
$T_{9A}$ is the tension of the Type IIA D9-brane.  

As before, the tachyon potential is not known exactly, but it
is known to have a reflection symmetry and the height of the
potential follows from the identification of the global minimum
with the closed string vacuum. We choose the global minimum to
be at $t=0$ for easier comparison with our previous results.
The reflection symmetry then acts by $(t - t_*) \rightarrow
-(t-t_*)$. 
A schematic potential with these properties is given in fig. 2.
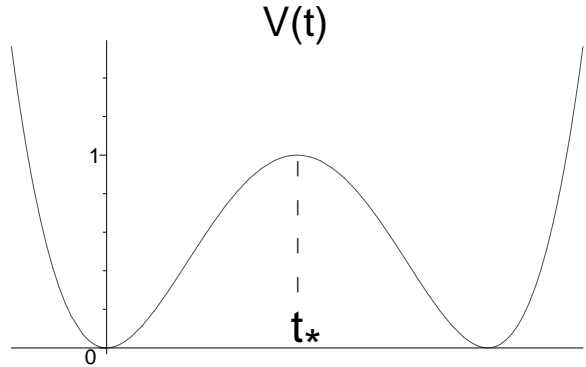
\fig{The open superstring tachyon potential.}{supertach.ps}{3truein}

Repeating the analysis done in the bosonic string shows that
the solution
\eqn\sfourteen{t= t_*\phi_0}
has the same tension as a D7-brane. By turning on constant 
$B$-fields in an even number of dimensions we construct the
rest of the non-BPS branes of IIA. Note that in contrast to
the bosonic string, here we obtain all non-BPS D-branes 
through this construction since in type II theory the
codimension of these branes differs by an even integer.

This solution is unstable as in the previous analysis
with the instability reflecting the presence of a tachyon
of the correct mass on the D7-brane. Multiple D-branes can
be incorporated as before, and the analysis of gauge field
couplings is essentially the same as the discussion in the
previous section. 
 
We can apply the reflection symmetry to obtain the reflected
solution \foot{Note that this solution is not of the canonical form
described in \gms\ of a critical point of the potential
times a projection operator. If $V'(t)=t \prod_i (t-\lambda_i)$
then $t= \lambda_k(P+1)$ with $P^2=P$ is a solution if $\lambda_k=
\lambda_i/2$ for some $i$.}
\eqn\sfref{t = t_* (2-\phi_0)~.}
This reflected solution represents a physically distinct configuration;
it is asymptotic to a solution whose ten-form RR field
strength differs by one unit from \sfourteen~\hhk.

We can also construct an unexpected solution (and its $Z_2$ image)
since the potential
has an additional stationary point at $t = 2 t_*$:
\eqn\spuzzle{t = 2 t_* \phi_0~.}
Since $V(2 t_*)$ vanishes, this soliton has vanishing tension
in the limit of infinite non-commutativity! It is easy
to see from the considerations in \gms\ that this object
is stable.  If we apply this 
construction in $10$ Euclidean dimensions we would seem to
obtain a zero action instanton. The presence of such an object
would clearly have dramatic consequences. 

One way to understand the solution \spuzzle\ is to consider
a level $k$ solution
\eqn\puzzlek{t_k = 2 t_* (\phi_0+\phi_1 + \ldots + \phi_{k-1})}
in the limit $k \rightarrow \infty$ as in the discussion in
section 2.4. This gives the solution $t_{\infty} = 2t_*$
which is simply the $Z_2$ image of the closed string vacuum
at $t=0$. In the same way that adding up an infinite number
of D-23 branes gave a D25-brane, here adding up an infinite number
of these tensionless 7-branes gives the other closed string vacuum.

It may well be
that at finite $\alpha' B$ this solution develops 
a non-zero tension representing the fact that
once derivatives are included it costs non-zero action to move
from one vacuum to the other. 
A subleading 
tension of order $1/(g_s \alpha' B)$ would also have interesting
consequences since there would then  be a one-parameter family of
non-perturbative objects  with variable tension. Since asymptotically one is 
in the closed
string vacuum and $B$ can be gauged to zero, the value of $B$ is not
a closed string modulus, but rather some modulus of the solution. 
In principle the tension could
be non-zero only at one string loop, but this seems unlikely since
it would modify 
the perturbative structure of string theory.  More work is
clearly needed to understand this mysterious object. 

The other unstable D-brane system of interest in type II string
theory is a $Dp - \overline{D p}$ system of BPS $Dp$-branes. These
objects have a complex tachyon with a ``Mexican hat'' potential
given by rotating fig.2 about a vertical axis at $t=t_*$. There are also
two $U(1)$ gauge fields, $A^+$ and $A^-$, coming from the
D-brane and anti D-brane respectively, and the tachyon carries
charge one under the relative $U(1)$ with gauge field $A^+ - A^- $. 
Both of the $U(1)$'s become non-commutative in the presence of
a non-zero $B$-field with the tachyon transforming in the
bi-fundamental representation. 

For topologically trivial solutions we
can use the relative $U(1)$ to make the tachyon field $t$ real.
The above solutions are then also solutions to this system. Since $t$
is real, the tachyon configuration does not act as a source for
the relative $U(1)$ gauge field and so the solutions carry no
net lower D-brane charge.  The
coefficient in front of the action is now $2 T_p$, 
so we can interpret the analogue of the first solution 
\sfourteen\ as  an unstable  $D(p-2) - \overline{D (p-2)}$ brane system
with energy $2 T_{p-2}$. The analogue of the second solution 
\spuzzle\ is apparently a
tensionless bound state of the  $D(p-2) - \overline{D(p-2)}$ system.

BPS $D(p-2)$-branes can be constructed as vortices in the complex
tachyon field \sena.  One can find an exact vortex solution 
using a  two-dimensional effective action which includes the
tachyon field and the $U(1) \times U(1)$ gauge fields\foot{The
previous version of this paper had a crucial minus sign error in the
equations of motion,
the vortex solution presented in \edtach\ was an important guide
in correcting this error.} 
%
%
%
%
%
%
%
%

The action is
\eqn\voractz{ S = \int dz d \bar z \left( \overline{D_\mu t} D^\mu t - 
{1 \over 4}
F^+_{\mu \nu}F^{+ \mu \nu} - {1 \over 4}
F^-_{\mu \nu}F^{- \mu \nu} - V(t, \bar t) \right) \ .}
The gauge field strengths have the canonical form and the covariant
derivatives are  given by
\eqn\coactz{D_\mu t = \partial_\mu t + i ( A^+_\mu t - t A^-_\mu) }
and
\eqn\coactzb{\overline{D_\mu T} = \partial_\mu \bar t -i (
\bar t A^+_\mu - A^-_\mu \bar t ). }

In the limit of large non-commutativity, where we can ignore derivative
terms, the  equations of motion are
\eqn\vorteoms{
   \eqalign{ [A^{+ \nu},[A^+_\nu,A^+_\mu]] & = 
  		A^+_\mu t \bar t - t A^-_\mu \bar t 
		+ t \bar t A^+_\mu - t A^-_\mu \bar t \cr
	[A^{- \nu},[A^-_\nu,A^-_\mu]] & = 
		A^-_\mu \bar t  t -\bar  t A^+_\mu  t + 
		\bar t  t A^-_\mu -\bar  t A^+_\mu  t \cr
	-A^+_\mu A^{+\mu} t+ 2 A^{\mu +} t A^-_\mu -
		t A^{- \mu}A^-_\mu  & = - \alpha tV' - \beta V' t\ . \cr 
}}
The values of $\alpha,\beta$ depend on the ordering prescription
for $V$. 

These equations are solved by 
\eqn\vortex{\eqalign{
	t &= t_* \sum_{i=0}^\infty \phi_{i,i+1} \cr 
A^+_z & = A^+_{\bar z}= t \bar t = 1 \cr
A^-_z & = A^-_{\bar z}=\bar t t = 1 - \phi_0 \  .\cr }}
Note that the field strengths of both
gauge fields vanish since $A^{\pm}_z$ are Hermitian and so
$F_{z \bar z} \sim [A_z,A_{\bar z}]=0$. Similarly, $D_\mu D^\mu t$ vanishes,
so to get a solution we can  choose the ordering in $V$
to be $V(\bar t t -1)$.  This gives $\alpha=1,\beta=0$ and
we  have a solution using $t V' \propto t  \phi_0=0$.

In the commutative case the vortex carries a non-zero D7 charge
which arises from the coupling of the RR potential to the
relative $U(1)$ gauge field strength,
$\int C_8 \wedge (F^+ - F^-)$.
In the large non-commutativity limit we have found $F^+-F^-=0$, so one
would naively think that there is no induced $D7$ charge.
This is incorrect. 
The couplings of RR fields to gauge and 
tachyon fields take an elegant form worked out in \kw. 
The relevant contribution is
\eqn\kweq{ \int C \wedge d\, {\Tr}\left(t\wedge \overline{Dt}\right)\ .
}
Substituting \vortex\ one finds that 
the last term gives the correct induced $D7$ charge. 
To show this, it is important to keep the spatial derivative term in
$D_\mu t$.


\newsec{Fundamental Strings as Electric Flux Tubes}

After tachyon condensation the open string degrees of freedom are confined 
and excitations of the theory are closed strings. We would like to describe  
these excitations as solitons. To this end we construct in  
the following an electric 
flux tube with the tension of the fundamental string. 
The classical confinement of electric flux found here is similar 
in spirit to that discussed in \ks, but distinct from that 
discussed in \refs{\yi,\bhy}. 
 
Naively, a gauge theory action of the type \tachgauge\ gives, 
in the strong coupling limit $h(t)G_s^{-1}\rightarrow 0$, 
very heavy electric flux tubes --- one expects a tension 
of order $h^{-1}G_s$.  However, with the nonlinearity inherent 
in the Born-Infeld action, we will see that 
the energy cost of a flux quantum 
saturates, and the flux tube remains light in the limit. 
 
\subsec{The flux tube and its tension} 
 
We need the effective action for the tachyon and the gauge fields for  
configurations carrying electric flux in a particular direction, say  
$x^1$. The task is simplified with the introduction of a large $B$-field  
in all $24$ transverse directions $x^i$, $i=2,\cdots,25$. Then  
derivatives  
along these directions are negligible. Furthermore, it is sufficient  
(for now) to consider gauge  
field configurations that are constant in time and along the flux tube. 
The tachyon potential in the presence of {\it constant} background 
fields is known from a theorem of Sen~\refs{\senact,\senuniv}:  
The potential is universal 
up to tachyon independent deformations of the overall metric. Thus, 
in our context, the action is of the Born-Infeld type 
\eqn\twtwo{ 
S = - \int d^{26}x~ V(t)\sqrt{-\det[\eta_{\mu\nu} + \pa F_{\mu\nu}]}~. 
} 
(Other recent discussions of the action include \refs{\bergs,\kluson}). 
Coordinates were chosen so $G_{\mu\nu}=\eta_{\mu\nu}$, and $\star$ products  
are implied. In this 
section we absorb the overall tension of the D25-brane in the potential  
$V(t_*)$.  
It is sufficient to retain just one component of the gauge field, {\it i.e.} 
\eqn\redact{ 
S = - \int d^{26}x ~V(t) \sqrt{1 - (2\pi\alpha^\prime {\dot A}_1)^2}~. 
} 
The analysis will be simplest in the canonical formalism. 
We therefore compute the momentum conjugate to the gauge  
field 
\eqn\twthree{  
{\cal E} = {V(t) (\pa)^2 {\dot A}_1 \over \sqrt{1 -  
(2\pi\alpha' {\dot A}_1)^2}}~. 
} 
Quantization of the electric flux ${\cal E}$ plays an important role in our
discussion.  In the large $B$ limit ${\cal E}$ is the electric flux of a 
$1+1$ dimensional $U(\infty)$ gauge theory.  Flux quantization in $1+1$
Yang-Mills theory is analyzed in \wittenflux, and we can apply a similar
analysis here.  Electric flux receives contributions from two sources:
from matter and gauge field charge fluctuations, and from charges at infinity.
In the present context these charges have values corresponding to the endpoints
of  open strings.  It follows that the electric flux in each $U(1)$ subgroup 
of $U(\infty)$ is quantized, and that it can be thought of roughly as the 
number of open strings (signed depending on orientation) ending on the
corresponding D23-brane.  More precisely, in a diagonal basis the eigenstates
of ${\cal E}$ are
\eqn\fluxq{
{\cal E} = {1 \over (2 \pi \theta)^{12}}
\sum_{k=0}^\infty n_k \phi_k~, \quad\quad n_k \in {\bf Z}.}
From now on we will drop the overall factors of $(2 \pi \theta)^{-12}$; they 
can be absorbed into field redefinitions. 
The overall $U(1)$ flux,
\eqn\totflux{
 \int\! d^{24}x \, {\cal E} = \sum_{k=0}^\infty n_k 
 \equiv N}
is a conserved quantity and can be identified with the total number of 
fundamental strings in the state.  The individual eigenvalues $n_k$ are
not conserved, and the general quantum state will correspond to a superposition
of different $n_k$.  In a complete quantum mechanical treatment it would be 
important to include the effect of transitions 
among different flux states,\foot{One might expect such fluctuations
to be rather large, since the effective coupling 
$1/V(t)$ in \redact\ (we have absorbed a factor of $1/g_s$
in the tachyon potential) is quite large as the tachyon field approaches
the closed string vacuum $V(t)\rightarrow 0$.  It would be interesting
if the fluctuations could be related to the usual divergent
vacuum fluctuations in the spatial position
of a perturbative closed string.
We thank S. Shenker and A. Lawrence for discussions on this point.}
but we will be less ambitious and consider the properties of a single flux
state of the form \fluxq.

We consider the Hamiltonian relevant to finding  
the configuration which minimizes the energy in a given total flux sector,
\eqn\hamlam{ 
H = \int d^{25}x \left[\sqrt{V(t)^2+ {\cal E}^2 /(\pa)^2 }  
+\lambda \partial_1{\cal E} \right] -\lambda N~, 
} 
where $\lambda$ is a Lagrange multiplier enforcing the condition that 
we consider configurations with $N$ flux quanta. The corresponding  
equations of motion are 
\eqn\eomone{ 
{\delta H\over\delta t} 
={V(t)V^\prime(t)\over\sqrt{V(t)^2+{\cal E}^2/(\pa)^2}}= 0~, 
} 
\eqn\eomtwo{ 
{\delta H\over\delta {\cal E}}= 
\left[{{\cal E}\over (\pa)^2\sqrt{V(t)^2 +  
{\cal E}^2/(\pa)^2}}+\lambda\right]=0~. 
} 
We want solutions that  
exist in the vacuum after tachyon condensation, so $V(t)$ should vanish  
asymptotically. We therefore try solutions of the form 
\eqn\toeo{ 
t = t_0\phi, 
} 
where $\phi$ is one of the functions satisfying $\phi\star\phi =\phi$.
\eomone\ is solved by taking $t_0$ to be an extremum of $V$.  
Taking $t=t_*$, \eomtwo\ is solved by 
\eqn\Ea{ 
{\cal E}  =  e\phi, 
}
and the flux quantization condition \totflux\ determines $e = N/k$ for $\phi$
at level $k$.   
\eomtwo\ also determines the Lagrange multiplier $\lambda$, but this is  
not needed in the subsequent discussion.  Inserting the solution back into 
the Hamiltonian \hamlam\ the tension is found to be
\eqn\pqener{ 
T= \sqrt{k^2 V(t_*)^2 + N^2 /(\pa)^2} ={1\over\pa}\sqrt{k^2/g^2_s + N^2}~. 
} 
This is as expected for an $(N,k)$ type string. 

An even simpler solution is found by taking $t=0$.  In this case
$V(0)=0$ so \eomtwo\ places no constraint on the form of ${\cal E}$, although
it does determine $\lambda$. ${\cal E}$ can take any form \fluxq\
consistent with the flux quantization condition \totflux.   The tension
of such a solution is 
\eqn\fsener{ 
T = \sqrt{V(0)^2+e^2/(\pa)^2}={N\over\pa}~, 
} 
which is the result expected for $N$ fundamental strings.

%
 
The way we derived it, the tension was essentially guaranteed to come  
out right. The nontrivial part was the magic of non-commutativity,  
which allows one to find a {\it localized} solution to equations which  
would otherwise be difficult to analyze. 
 
The action that we started with \twtwo\ is proportional to $1/g_s$ (absorbed 
in $V(t)$). As mentioned above, 
it is therefore surprising that the flux tube solution \fsener\ has energy  
of ${\cal O}(g_s^0)$.  
This is possible because, in the nonlinear Born-Infeld 
Hamiltonian \hamlam, the coefficient of the electric field 
term is independent of $V(t)$, whereas it would have  
a coefficient $1/V$ in a Yang-Mills type action. 
Alternatively, the usual expansion of the Born-Infeld action 
in powers of the field strength is inappropriate; 
one is in a `relativistic' limit of the field dynamics, 
with $V$ playing the role of mass. 
 
\subsec{Fluctuations} 
We would like to consider also excitations of the fundamental string. 
This is more demanding because generally the action \twtwo\  
receives corrections with unsuppressed  
derivatives in the longitudinal direction $x^1$.  
However, the action is still applicable for fluctuations with large  
wavelength $\sqrt{\alpha^\prime}F^\prime_{\mu\nu}\ll F_{\mu\nu}$ 
(prime denotes derivatives along $x^1$).  Note  
that this condition still allows for amplitudes of order string scale  
$\alpha^\prime F_{\mu\nu}\sim 1$ so the nonlinear terms in the 
Born-Infeld action do contain valid information.   As mentioned earlier,
a true quantum state corresponding to a fundamental consists of a 
superposition of solutions with a given total electric flux.  However, we will
consider the simpler exercise of analyzing  fluctuations around a single 
solution with fixed $n_k$, 
and leave the more complete treatment for future work.  
 
Again, our considerations will be simplest in the canonical formalism. 
To find the Hamiltonian, we expand the determinant  
in \twtwo\ and write the action 
\eqn\lagcan{ 
S =  - \int d^{26}x~ V(t)\sqrt{\det({\One}+{\bf F}) 
(1 - F_{0\alpha}M^{\alpha\beta}F_{0\beta})}~, 
} 
where 
\eqn\modmet{M^{\alpha\beta} = \left( {{\One}\over  
{\One}+{\bf F}}\right)^{\alpha\beta}_{\rm sym}~. 
} 
Here $\alpha,\beta,\cdots$ are purely spatial indices and the matrices  
${\One},{\bf F}$ have components $\delta_{\alpha\beta},F_{\alpha\beta}$.  
In this subsection we take $\pa=1$ to simplify the formulae; these 
factors are easily restored by comparing with the previous computation. 
The  canonical momenta are 
\eqn\canmo{ 
{\cal E}^\alpha =  
{V(t)M^{\alpha\beta}F_{0\beta}\over\sqrt{1 - F_{0\alpha}M^{\alpha\beta} 
F_{0\beta}}} ~, 
} 
and the Hamiltonian becomes  
\eqn\hamil{ 
H = \int d^{25}x~\left({\cal E}^\alpha {\dot A}_\alpha - {\cal L}\right) = 
\int d^{25}x~\left[ \sqrt{{\cal E}^\alpha M_{\alpha\beta} {\cal E}^\beta + 
V(t)^2 \det({\One}+{\bf F})} +  
A_0\nabla_\alpha {\cal E}^\alpha\right] ~, 
} 
where 
\eqn\met{ 
M_{\alpha\beta} = \delta_{\alpha\beta} - F_{\alpha\gamma}F^\gamma_{~\beta} 
~.} 
The equation of motion for $A_0$ is the Gauss law constraint 
\eqn\gauss{ \nabla_\alpha {\cal E}^\alpha =0~. 
} 
Hereafter it is convenient to choose the gauge $A_0=0$. 
 
Derivatives in the transverse directions are negligible, so transverse  
field strengths simplify: $F_{ij}=0$ and $F_{1i}=A^\prime_i$.  
The Hamiltonian can therefore  
be written 
\eqn\hamtwo{ 
H = \int d^{25}x \sqrt{({\cal E}^1)^2  
( 1 + (\vec{A}^\prime)^2) + {\vec{\cal E}}^2 +  
({\vec{\cal E}}\cdot {\vec A}^\prime)^2 + {V(t)^2 (1+(\vec{A}^\prime)^2)}} 
~. 
} 
The vector notation applies to the transverse directions $x^i$,  
$i=2,\cdots,25$. 
Considering just the fundamental string we can take $V(t)=0$ from here 
onwards. 
 
Fluctuating strings are described as flux tube solutions based on functions  
satisfying $\phi\star\phi =\phi$, as before, but with the origin displaced  
in a manner depending on the coordinates along the string 
\eqn\fsans{ 
\phi = \phi(x^i - f^i (x^0,x^1))~. 
} 
A suitable {\it ansatz} is 
\eqn\fansatz{  
{\cal E}_1 = e_1\phi, \quad 
{\vec{\cal E}}={\vec e}\phi, \quad 
{\vec A}^\prime = {\vec a}^\prime \phi~. 
}   
The coefficient $e_1$ will generally be $N/k$,
but in this section we consider 
a single string, so $e_1=1$.  
The functions ${\vec e},{\vec a}$ depend on the string  
coordinates $x^a=x^0,x^1$ but not on the transverse coordinates $x^i$. 
 
We want to find the effective dynamics of the variables $f^i$.  
It is not sufficient to insert our 
{\it ansatz} in the Hamiltonian \hamtwo\ because that would 
not ensure that the equations of motion  
for other fields are satisfied.  
The correct procedure is known as Hamiltonian reduction (it is described  
in detail in a closely related context in \lwone). 
By solving the constraints,
the Hamiltonian is expressed in terms of the reduced  
variables ${\vec f}$ and their conjugate momenta.
In the present context the important constraint is 
Gauss' law \gauss. Inserting our {\it ansatz} we find 
\eqn\gaussone{ 
-e_1 {\partial\phi\over\partial x^i}{\partial f^i\over\partial x^1}+ e^i 
{\partial\phi\over\partial x^i}=0 
~.} 
Recalling that $e_1=1$, a simple solution is 
\eqn\gausstwo{  
{\vec f}^\prime = {\vec e}~. 
} 
We need to find the momentum conjugate to ${\vec f}^\prime$. The fields  
${\vec{\cal E}}$ and ${\vec A}$  
are canonically conjugate in the original theory.  
This implies that ${\vec e}$ and ${\vec a}$ are  
canonically conjugate in the reduced  
theory. From \gausstwo\ we therefore find that the momenta conjugate  
to ${\vec f}^\prime$ are $\vec{\pi}\equiv {\vec a}^\prime$.\foot{Up to 
a factor of the wavenumber, when we expand in the usual oscillator basis.} 
We conclude that the reduced Hamiltonian  
\eqn\redham{ 
H = \int dx^1~ \sqrt{1+{\vec\pi}^2 + ({\vec f}^\prime)^2 
+ (\vec{\pi}\cdot \vec{f}^\prime )^2}~, 
} 
describes the fluctuations of the flux tube. The dynamics of  
transverse modes was ``integrated out'' in the Hamiltonian reduction.  
 
The significance of this Hamiltonian is best exhibited in the Lagrangian 
formalism. We therefore compute the time derivative using Hamilton's 
equations 
\eqn\fdot{ 
{\dot{\vec f}} = {\delta H\over \delta {\vec e}} 
= { \vec{\pi}+ {\vec f}^\prime ({\vec f}^\prime \cdot {\vec\pi}) \over 
\sqrt{1+{\vec\pi}^2 + ({\vec f}^\prime)^2 
+ ({\vec\pi}\cdot \vec{f}^\prime )^2}} 
~,} 
and find 
\eqn\redlag{ 
L = \int d^2 x ~{\vec\pi}\cdot{\dot{\vec f}}  -\int dx^0 ~H 
=- \int d^2 x \sqrt{ (1-(\dot{\vec f})^2)(1+(\vec{f}^\prime)^2) 
+ ({\dot{\vec f}}\cdot {\vec f}^\prime)^2} 
~.} 
This should be compared with the standard Nambu-Goto action  
\eqn\ngact{ 
L_{NG} = - \int d^2 x \sqrt{-{\rm det}[\partial_a X_\mu \partial^a X^\mu]} 
= -\int d^2 x \sqrt{ ({\dot{\vec X}})^2 ({\vec X}^\prime)^2 -  
({\dot{\vec X}}\cdot {\vec X}^\prime)^2} 
~.} 
In the static gauge $X^\mu = (x^0,x^1, f^i)$ used here this reduces  
precisely to \redlag. The effective action for the fluctuations 
of the flux tube is therefore the Nambu-Goto action! 
 
The next step is to quantize the fluctuations of the flux tube. Their 
action is the Nambu-Goto action so the result is 
that of a closed fundamental string.  
We have justified the Lagrangian \twtwo\  
only for large semi-classical waves so it is in this regime  
that the spectrum of the flux tube should be compared with that of  
fundamental strings. Our computation demonstrates a perfect agreement  
in the allowed energies and their degeneracies. 
 
It is tantalizing that, if we extend this reasoning beyond its apparent  
regime of validity, we find the entire fundamental string spectrum as  
simple excitations in the open string field theory. This would include  
{\it very} light modes --- such as the massless graviton, and even the closed  
string tachyon. Usually this kind of identification would be impossible  
{\it in principle}: the quantum fluctuations of a soliton are collective  
excitations rather than fundamental objects, because the soliton  
itself is made out of more basic constituents.  
We are better off here because, after tachyon condensation, the flux tubes  
are the lightest  
objects in the theory and therefore subject to quantization. Although 
this removes the objection of principle, we presently have no  
justification to trust our result for light modes. 
 
\subsec{Multiple strings and interactions} 
 
One can also consider electric fluxes lying in $U(k)$ subgroups
of the $U(\infty)$ gauge group, leading to additional solutions. 
These parallel closely the construction of Matrix string theory \dvv.
With $x^1$ compactified on a large circle of radius $R$,
one can describe multiply wrapped strings via the holonomy
of the gauge field that twists the eigenvalues of the
$U(k)$ electric field into a `long string' or `slinky'.
One can follow the arguments given in \dvv\ leading
to the identification of the interactions of such a string with
the effective twist operator that arises when
$SU(2)$ symmetry is restored by two strands of
the Matrix string coming together in the transverse space.
Thus we see that the flux tubes at least qualitatively
interact in the proper way; it would be interesting to
see if at least some perturbative string amplitudes can
be reproduced within the non-commutative framework 
(for instance, tree level amplitudes that just exchange winding 
among these macroscopic strings).  A potential difficulty
in this exercise is that the original string coupling constant
$g_s$ appears only through the tachyon potential $V(t)$ in \hamtwo;
after tachyon condensation $V\rightarrow 0$, thus it seems
that the electric flux lines do not know about the original
string coupling.  
One does not a priori know why the interaction strength of
the flux lines is $\CO(g_s)$ and not $\CO(1)$.

One might also be concerned that there appear to be two separate
descriptions of a $k$ times wound fundamental string:
Namely $k$ units of flux in a $U(1)$ solution, 
as well as $k$ units of flux wound into a `slinky' in $U(k)$.
In Matrix string theory, these two carry different quantum numbers;
the rank of the gauge group corresponds to the number of
bits of discrete light cone momentum of the Matrix string,
while the electric flux represents D0-brane charge that the
string is bound to.  In the present context, 
both of these objects are embedded in the same $U(\infty)$ gauge theory, 
and simply represent different spatial distributions of the
flux lines; there is no apparent reason why the 
the path integral over the configuration space of the flux lines
in the strongly coupled gauge theory
will not smoothly evolve these two flux configurations into one another.  

To summarize, we find macroscopic fundamental strings appearing
as light electric flux excitations of the open string field theory, after
the tachyon is condensed to form the closed string vacuum.
The tension of the flux tubes is just right, and the
fluxes join and split in the manner familiar from
string perturbation theory (although it is not at present understood 
whether their interaction strength
is related to the string coupling $g_s$).

 
\newsec{Comments and questions} 
 
It is remarkable that turning on a large $B$-field simplifies 
the structure of open string field theory in such a dramatic 
way that one can construct exact D-brane and fundamental string 
solutions. The full force of this construction relies heavily 
on Sen's conjecture that tachyon condensation represents the 
closed string vacuum. We use this conjecture to determine the 
height of the tachyon potential, but also to argue that at the 
end of the construction one can gauge $B$ to zero far from 
the solution so that one is discussing D-branes and fundamental 
string in the usual closed string vacuum with $B=0$. 

Similarities to Matrix theory have been a persistent thread 
running through our discussion.  In both cases
a large dimensionless parameter is introduced ---
$\alpha' B$ for the non-commutative limit, and the boost rapidity
in Matrix theory --- upon which quantities of interest do not depend,
yet whose introduction facilitates calculations.
The large $B$-field of non-commutative geometry
induces a macroscopic density of lower-dimensional D-branes,
while tachyon condensation removes the higher-dimensional
D-brane and makes the effective gauge coupling large in regions 
described by the closed string vacuum.  
Thus the ingredients of the Matrix theory limit are present;
furthermore, the scaling limit of \sw\ is the analogue of the
Maldacena scaling limit \malda\ ($\alpha'\rightarrow 0$,
with the energies of stretched open strings held fixed).
It would be interesting to make the relation
between these two circles of ideas more precise.
 
Our work raises a number of other interesting questions: 
 
\item{1.} Is it possible to also construct the NS 5-brane (or 
21-brane in the bosonic string) using these methods\foot{The analogy 
with Matrix theory suggests that the problem
is similar in nature to the construction of the 
transverse Matrix fivebrane.   In particular,
the energy density of the object scales as $g_s^{-2}$,
whereas classical open string field configurations have energy 
densities scaling as $g_s^{-1}$.}? 
\item{2.} How does one understand the freezing out of the 
open string degrees of freedom and the ability to gauge $B$ 
to zero in the closed string vacuum, directly in string field 
theory? 
\item{3.} What is the interpretation of the very light solitons 
(massless to leading order in $1/\alpha'B$)
found in the superstring? 
\item{4.} What are the leading $1/\alpha' B$ corrections 
to the results presented here? 
\item{5.} What is the coupling strength of the fundamental string 
constructed in section 4? 
 
\noindent
We hope to address some of these questions in future work.

\bigskip\medskip\noindent 
{\bf Acknowledgements:} 
We would like to thank  
Rajesh Gopakumar, 
Kentaro Hori, 
Shamit Kachru, 
Albion Lawrence,
Shiraz Minwalla, 
Steve Shenker, 
Eva Silverstein,  
and 
Andy Strominger 
for helpful conversations.  
This work was supported by DOE grant DE-FG02-90ER-40560 
and NSF grant PHY-9901194.  
F.L. was supported in part by a Robert R. McCormick fellowship. 
P.K. thanks the Harvard and Stanford theory groups for hospitality; 
J.H. would like to thank the Institute for Advanced Study for hospitality; 
and 
E.M. thanks the Rutgers University theory group for hospitality.

\bigskip\medskip\noindent
{\bf Note Added:}
As this work was being typed we became aware of work by
K. Dasgupta, S. Mukhi, and G. Rajesh \dmr\
which overlaps our discussion of
non-commutative solitons in type II string theory. We thank them
for bringing this work to our attention. 
 
\listrefs 
\end